\documentclass[conference]{IEEEtran}
\IEEEoverridecommandlockouts
\usepackage{cite}
\usepackage{amsmath,amssymb,amsfonts}
\usepackage{algorithmic}
\usepackage{graphicx}
\usepackage{textcomp}
\usepackage{xcolor}
\usepackage{adjustbox}
\usepackage{multirow}
\usepackage{caption}
\usepackage{booktabs}
\usepackage{dsfont}
\def\BibTeX{{\rm B\kern-.05em{\sc i\kern-.025em b}\kern-.08em
    T\kern-.1667em\lower.7ex\hbox{E}\kern-.125emX}}
\begin{document}

\title{StarTSE: Towards Streaming Target Speaker Extraction via Chunk-wise Interleaved Splicing of Autoregressive Language Model}
\author{
    \IEEEauthorblockN{Shuhai Peng\textsuperscript{1,*}, Hui Lu\textsuperscript{2,*}, Jinjiang Liu\textsuperscript{1}, Liyang Chen\textsuperscript{1}, Guiping Zhong\textsuperscript{3}, Jiakui Li\textsuperscript{3}\\
    Huimeng Wang\textsuperscript{2}, Haiyun Li\textsuperscript{1}, Liang Cao\textsuperscript{1}, Shiyin Kang\textsuperscript{3}, Zhiyong Wu\textsuperscript{1,\textdagger}}
    \IEEEauthorblockA{
        \textsuperscript{1}Shenzhen International Graduate School, Tsinghua University \\
        \textsuperscript{2}The Chinese University of Hong Kong \quad \textsuperscript{3}SenseTime Research \\
        \textsuperscript{*}Equal contribution, \textsuperscript{\textdagger}Corresponding author
    }
}

\maketitle

\begin{abstract}
While generative models have set new benchmarks for Target Speaker Extraction (TSE), their inherent reliance on global context precludes deployment in real-time applications. Direct adaptation to streaming scenarios often leads to catastrophic inference performance degradation due to the severe mismatch between training and streaming inference. To bridge this gap, we present the first autoregressive (AR) models tailored for streaming TSE. Our approach introduces a Chunk-wise Interleaved Splicing Paradigm that ensures highly efficient and stable streaming inference. To ensure the coherence between the extracted speech segments, we design a historical context refinement mechanism that mitigates boundary discontinuities by leveraging historical information. Experiments on Libri2Mix show that while AR generative baseline exhibits performance degradation at low latencies, our approach maintains 100\% stability and superior intelligibility. Furthermore, our streaming results are comparable to or even surpass offline baselines. Additionally, our model achieves a Real-Time-Factor (RTF) of 0.248 on consumer-level GPUs. This work provides empirical evidence that AR generative backbones are viable for latency-sensitive applications through the Chunk-wise Interleaved Splicing Paradigm.
\end{abstract}

\begin{figure*}[ht]
    \centering

    \includegraphics[width=0.75\textwidth]{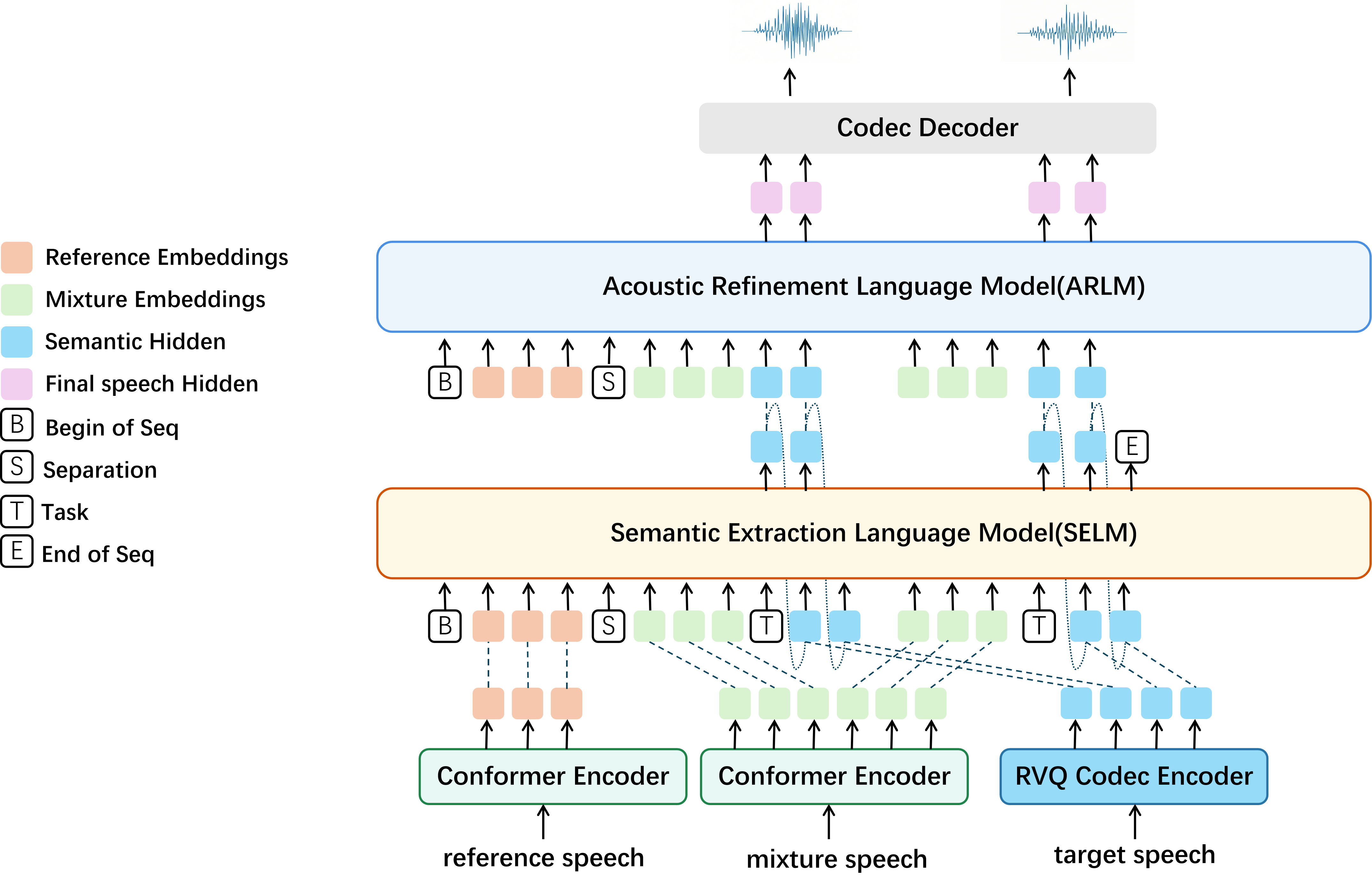} 
    
    \caption{Overview of our framework. The input mixture is processed as a sequence of discrete chunks to ensure strict causality.}
    \label{fig:arch}
    
\end{figure*}


\section{Introduction}
\label{sec:intro}

Target Speaker Extraction (TSE)\cite{b6} aims to isolate the speech of a specific speaker from a complex acoustic mixture containing interfering speakers and background noise. Unlike Blind Source Separation\cite{b6}, which segregates all concurrent audio sources without distinction, TSE exploits auxiliary reference cues, typically a pre-recorded enrollment utterance of the target speaker. This capability could benefit real-time scenarios to capture target input accurately, such as teleconferencing systems, voice-controlled assistants and multi-turn dialogue interactions.

Historically, the TSE domain has been dominated by discriminative approaches, such as SpEx+\cite{b7} and VoiceFilter-Lite\cite{b2}, which rely on estimating masks or filters to suppress interference. While computationally efficient, these methods often introduce processing artifacts and struggle to reconstruct missing high-frequency spectral details. 
Recently, the paradigm has shifted towards generative modeling for superior speech quality. Making valuable contributions to the field, innovative architectures such as DPM-TSE\cite{hai2023dpmtsediffusionprobabilisticmodel} and Diff-TSE\cite{kamo2023targetspeechextractionconditional} harness diffusion probabilistic modeling, while TSELM-L\cite{tang2024tselmtargetspeakerextraction} and LauraTSE\cite{b5} successfully exploit the power of Transformer architecture\cite{vaswani2023attentionneed}. Recently, SoloAudio\cite{wang2025soloaudiotargetsoundextraction} and SoloSpeech\cite{wang2025solospeechenhancingintelligibilityquality} have introduced Diffusion Transformers to the field. By treating speech extraction as a conditional generation task, these frameworks have achieved substantial improvements in quality and naturalness compared to traditional discriminative methods.

However, for real-time scenarios, fundamental misalignment persists between the global context dependency of current generative TSE models and streaming inference. Generative TSE models are predominantly designed for offline processing, relying heavily on global context dependency to capture long-term acoustic correlations and ensure high quality. Forcing such models to operate online streaming TSE task without future information inevitably leads to a severe training-inference mismatch, resulting in performance degradation. When adapting these generative frameworks to streaming scenario, two primary backbones emerge: Diffusion-based and autoregressive-based models, while Diffusion models offer superior generative diversity and robustness to noise, their iterative sampling nature imposes a significant computational bottleneck, making them inherently less efficient for chunk-wise streaming inference compared to the one-pass incremental execution of AR architectures. 

To address the misalignment, we present a novel streaming framework based on LauraGPT\cite{b3} backbone. We propose Chunk-wise Interleaved Splicing Paradigm, a mechanism designed to tackle the conflict between high-quality generation and real-time processing in traditional generative methods. Our approach reconfigures the input by interleaving mixture chunks and corresponding target codec representation. Specifically, we implement an interleaving strategy within both the Semantic Extraction Language Model (SELM) and Acoustic Refinement Language Model (ARLM) stages. This design enforces strict causal constraints, ensuring that computation at any step depends solely on historical and current information, preventing future context leakage. 

Our experimental results indicate that our streaming results are comparable to or even exceed those of the offline discriminative models Spex+\cite{b7} and WeSep\cite{wang2024wesepscalableflexibletoolkit}, in terms of speech quality and intelligibility. We achieved results comparable to or better than the generative model LauraTSE within 560ms or less. By introducing historical context refinement, both speech quality and speech intelligibility are improved. At the same time, the processing real-time factor (RTF) of 0.248 is achieved on consumer-level GPUs.

The main contributions of this paper are summarized 
as follows:

\begin{enumerate}
    \item First AR-based Streaming TSE Framework: To the best of our knowledge, we present the first autoregressive generative backbone tailored for streaming Target Speaker Extraction. This work provides empirical proof that generative models can be adapted to latency-sensitive scenarios, bridging the gap between high-quality generation and real-time processing.
    \item Chunk-wise Interleaved Splicing Paradigm: This interleaved paradigm supports native streaming speech processing. By employing append-only inference operation, it ensures constant computational complexity per chunk, significantly enhancing the training and inference efficiency of the autoregressive (AR) model.
    \item Better performance under low latency: Experiments demonstrate that our framework exhibits superior intelligibility and 100\% stability over AR generative baseline in low-latency scenarios. Notably, it delivers performance that matches or even exceeds offline baselines, providing a robust solution for real-time applications. Furthermore, we achieve a Real-Time Factor (RTF) of 0.248 on consumer-grade GPUs.
    \item Historical Context Refinement Mechanism: We design a mechanism to mitigate boundary discontinuities by leveraging historical information from previous chunk. This ensures inter-chunk phase and semantic coherence, significantly enhancing speech quality and intelligibility.
\end{enumerate}

\section{METHOD}

Target Speaker Extraction (TSE) aims to isolate the target speech from mixture guided by a reference speech. We propose a novel streaming adaptation framework based on the LauraGPT\cite{b3} backbone that employs a coarse-to-fine hierarchical strategy. As shown in Fig.~\ref{fig:arch}, the overall architecture consists of four key components: (A) Shared Conformer encoder extracts frame-level continuous embeddings from the mixture and reference. (B) Semantic Extraction Language Model (SELM) predicts the coarse-grained semantic codec tokens. (C) Acoustic Refinement Language Model (ARLM) captures the fine-grained acoustic details. (D) Codec decoder reconstructs the waveform from the latent representations.

\subsection{Feature Extraction}
The system processes mixture speech as a sequence of discrete chunks rather than full utterances to facilitate real-time streaming. Both the input mixture chunk and the reference speech are transformed into log-mel spectrograms and processed through two weight-shared, strictly causal conformer encoders to extract frame-level continuous embeddings $E_{\text{mix}}$ and $E_{\text{ref}}$. This design ensures that feature extraction depends exclusively on current and historical information, effectively preventing any future context leakage.

\subsection{Semantic Extraction Language Model (SELM)}

The Semantic Extraction Language Model forms the main part of our hierarchical architecture, responsible for capturing the semantic information. Generative TSE models typically concatenate the full reference, mixture, and target tokens (e.g., $[E_{\text{ref}}, E_{\text{mix}},U]$) as input to perform global attention. In streaming scenarios, this global attention mechanism relies on future context 
and results in severe inference performance degradation under low-latency constraints. To make it adapt to streaming scenarios, we propose the Chunk-wise Interleaved Splicing Paradigm.

We segment the mixture embedding $E_{\text{mix}}$ into chunks $\mathbf{C} = \{C_{\text{mix}}^{(1)}, \dots, C_{\text{mix}}^{(t)}\}$ and the target tokens $\mathbf{U} = \{u^{(1)}, \dots, u^{(T)}\}$. For each step $t$, we construct the input $\mathcal{S}_{\text{SELM}}^{(t)}$ by concatenating the static reference with the interleaved sequence of mixture chunks and target tokens:
\begin{equation}
\mathcal{S}_{SELM}^{(t)} = [ \underbrace{E_{\text{ref}}, v_{\text{sep}}}_{\text{Static Prefix}}, \underbrace{C_{\text{mix}}^{(1)}, v_{\text{task}}, u^{(1)}, \dots, C_{\text{mix}}^{(t)}, v_{\text{task}}, u^{(t)}}_{\text{Interleaved Sequence}} ]
\end{equation}
\begin{equation}
p(U \mid E_{\text{ref}}, C_{\text{mix}}) = \prod_{t=1}^{T} p(u^{(t)} \mid E_{\text{ref}}, C_{\text{mix}}^{(1:t)}, U_{\text{SELM}}^{(1:t-1)})
\end{equation}
where $v_{\text{sep}}$ and $v_{\text{task}}$ are special tokens for separation and task specification, and $U_{\text{SELM}}^{(1:t-1)}$ denotes the history of predicted semantic token chunks. By interleaving the mixture chunk $C_{\text{mix}}^{(t)}$ and its corresponding target tokens $u^{(t)}$, this paradigm enforces a hard temporal boundary. This guarantees causality by limiting the receptive field to historical observations, effectively preventing future leakage.
\subsection{Acoustic Refinement Language Model (ARLM)}

We introduce the Acoustic Refinement Language Model (ARLM) to recover the fine-grained acoustic information. This stage is crucial for recovering high-frequency details and improving the acoustic quality of the extracted speech.

Like SELM interleaved input strategy, the ARLM operates on the chunk-level interleaved inputs. To ensure strict causality while maintaining context coherence, we construct the input sequence for the ARLM denoted as $\mathcal{S}_{\text{ARLM}}$, is constructed by interleaving the sequence of mixture chunks with their corresponding predicted discrete tokens, appended to a static prefix containing the reference embedding. Formally, the input at step $t$ is formulated as:
\begin{equation}
\mathcal{S}_{\text{ARLM}}^{(t)} = [\underbrace{E_{\text{ref}}, v_{\text{sep}}}_{\text{Static Prefix}}, \underbrace{C_{\text{mix}}^{(1)}, U_{\text{SELM}}^{(1)}, \dots, C_{\text{mix}}^{(t)}, U_{\text{SELM}}^{(t)}}_{\text{Interleaved Sequence}}]
\end{equation}
\begin{equation}
p(H \mid E_{\text{ref}}, C_{\text{mix}}, U_{\text{SELM}}) = \prod_{t=1}^{T} p(h^{(t)} \mid E_{\text{ref}}, C_{\text{mix}}^{(1:t)}, U_{\text{SELM}}^{(1:t)})
\end{equation}
where $U_{\text{SELM}}^{(t)}$ represents the coarse-grained semantic tokens predicted by the SELM at step $t$, and $h^{(t)}$ denotes the ARLM's output of acoustically refined hidden states at step $t$.

\subsection{Codec Decoder}
A major challenge in chunk-wise streaming generation is boundary discontinuity, which often leads to lower speech quality. To address this, we introduce the Historical Context Refinement Mechanism at codec decoder stage.

Instead of generating each speech chunk in isolation, we explicitly leverage the output chunk from the previous step $t-1$ to guide the generation at step $t$. We concatenate the input of codec decoder as:
\begin{equation}
\text{Codec\ Input}^{(t)} = \operatorname{Concat}\left(h^{(t-1)}, h^{(t)}\right)
\end{equation}
By recirculating the refined hidden states back into the input stream, the model maintains a continuous phase and semantic context across chunk boundaries. This mechanism acts as a refiner, significantly enhancing the speech quality and intelligibility in low-latency scenarios.

\subsection{Training Objective}
To support the coarse-to-fine generation, the model is optimized end-to-end using a hybrid objective function consisting of the Negative Log-Likelihood ($\mathcal{L}_{\text{NLL}}$) for the semantic and the regression loss ($\mathcal{L}_{\text{REG}}$) for acoustic refinement:
\begin{equation}
\mathcal{L}_{\text{total}} = \lambda_1 \mathcal{L}_{\text{NLL}} + \lambda_2 \mathcal{L}_{\text{REG}}
\end{equation}
where $\lambda_1$ and $\lambda_2$ balance the semantic and acoustic reconstruction tasks.

\begin{table*}[t]
\centering
\caption{Performance comparison and streaming feasibility. Perceptual quality, intelligibility, speaker similarity, and Inference Success Rate (ISR) are evaluated across varying chunk sizes and offline. In the ``Category" column, ``G" represents generative models, and ``D" represents discriminative models. \textbf{Bold} denotes results superior to LauraTSE under the same chunk constraints.}
\label{tab:main_results}

\setlength{\tabcolsep}{3.0pt} 
\renewcommand{\arraystretch}{1.1}

\begin{tabular}{lllccccccccc}
\toprule
\multirow{2}{*}{\textbf{Method Setup}} & \multirow{2}{*}{\textbf{Category}} & \multirow{2}{*}{\textbf{Chunk Size}} & \multicolumn{3}{c}{\textbf{DNSMOS$\uparrow$}} & \multirow{2}{*}{\textbf{NISQA$\uparrow$}} & \multirow{2}{*}{\textbf{SpeechBERT$\uparrow$}} & \multirow{2}{*}{\textbf{WER}$\downarrow$} & \multicolumn{2}{c}{\textbf{Sim} $\uparrow$} & \multirow{2}{*}{\textbf{ISR$\uparrow$}} \\

\cmidrule(lr){4-6} \cmidrule(lr){10-11}
 & & & \textbf{SIG} & \textbf{BAK} & \textbf{OVL} & & &  & \textbf{WavLM} & \textbf{Wespeaker} &   \\
\midrule

Mixture & - & \textit{Offline} & 3.383 & 3.098 & 2.653 & 2.453 & 0.572 & 0.580  & 0.847 & 0.759 & 100.00\% \\
\midrule

Spex+\cite{b7} & D & \textit{Offline} & 3.472 & 4.027 & 3.186 & 3.349 & 0.878 & - & 0.973 & 0.935 & - \\

WeSep\cite{wang2024wesepscalableflexibletoolkit} & D &  \textit{Offline} & 3.486 & 3.838 & 3.118 & 3.892 & 0.895 & - & 0.980 & 0.945 & - \\
\midrule

TSELM-L\cite{tang2024tselmtargetspeakerextraction} & G &  \textit{Offline} & 3.489 & 4.041 & 3.212 & 3.961 & 0.793 & - & 0.887 & 0.627 & - \\
\midrule

\multirow{7}{*}{\shortstack[l]{LauraTSE\cite{b5}}} & \multirow{7}{*}{G}
 & \textit{Offline} & \textit{3.607} & \textit{4.078} & \textit{3.336} & \textit{4.330} & \textit{0.906} & \textit{0.082} &  \textit{0.973} & \textit{0.874} & \textit{100.00\%} \\
 & & 80ms & 1.254 & 1.120 & 1.105 & 1.031 & 0.294 & 0.960  & 0.522 & 0.504 & 15.07\% \\
 & & 160ms & 1.921 & 1.787 & 1.516 & 1.261 & 0.446 & 0.759 & 0.682 & 0.519 & 23.57\% \\
 & & 400ms & 3.325 & 3.660 & 2.921 & 2.899 & 0.787 & 0.249 & 0.931 & 0.575 & 89.37\% \\
 & & 560ms & 3.477 & 3.879 & 3.130 & 3.494 & 0.831 & 0.174 & 0.954 & 0.739 & 99.10\% \\
 & & 800ms & 3.537 & 3.976 & 3.224 & 3.859 & 0.859 & 0.123 & 0.964 & 0.804 & 99.73\% \\
 & & 2000ms & 3.592 & 4.057 & 3.310 & 4.266 & 0.895 & 0.090 & 0.972 & 0.843 & 99.93\% \\
\midrule

\multirow{6}{*}{\shortstack[l]{Proposed Streaming Method}} & \multirow{6}{*}{G}
 & 80ms & \textbf{3.025} & \textbf{1.898} & \textbf{1.896} & \textbf{1.295} & \textbf{0.692} & \textbf{0.300}  & \textbf{0.918} & \textbf{0.738} & \textbf{100.00\%} \\
 & & 160ms & \textbf{3.317} & \textbf{2.697} & \textbf{2.434} & \textbf{1.876} & \textbf{0.762} & \textbf{0.229}  & \textbf{0.941} & \textbf{0.797} & \textbf{100.00\%} \\
 & & 400ms & \textbf{3.490} & 3.606 & \textbf{3.010} & 2.891 & \textbf{0.830} & \textbf{0.152} & \textbf{0.958}  & \textbf{0.840} & \textbf{100.00\%} \\
 & & 560ms & \textbf{3.535} & 3.752 & 3.117 & 3.283 & \textbf{0.847} & \textbf{0.138} & \textbf{0.959}  & \textbf{0.847} & \textbf{100.00\%} \\
 & & 800ms & \textbf{3.559} & 3.923 & 3.222 & 3.593 & \textbf{0.863} & 0.133 & 0.961 & \textbf{0.847} & \textbf{100.00\%} \\
 & & 2000ms & 3.576 & 3.972 & 3.263 & 3.989 & 0.889 & 0.099 & 0.970 & \textbf{0.863} & \textbf{100.00\%} \\
\bottomrule
\end{tabular}
\vspace{-0.2cm}
\end{table*}

\section{EXPERIMENTAL SETUP}

\subsection{Dataset}
To ensure a fair comparison with the AR generative baseline and to benchmark against established offline systems, we adopted a unified data generation protocol based on LibriSpeech-460h\cite{7178964} and Libri2Mix\cite{cosentino2020librimixopensourcedatasetgeneralizable}. We followed the configuration in LauraTSE\cite{b5}, setting the mixing SNR between 0 and 5 dB and the reference speech duration to 5 seconds. This alignment ensures that our evaluation focuses on the efficacy of the proposed framework.

\subsection{Implementation Details} 
The generative backbone is the LauraGPT\cite{b3}. We utilize a pre-trained 16kHz funcodec\cite{b9} for getting the discrete acoustic token. Specifically, the model employs Residual Vector Quantization (RVQ) with 32 quantizers and a codebook size of 1024. The model contains approximately 89M parameters. To accelerate the training progress and enable large-batch optimization for stable convergence, the training was conducted on a distributed computing cluster comprising 13 nodes, each equipped with 8 NVIDIA V100-32GB GPUs (i.e., 104 GPUs).

\subsection{Evaluation Metrics} 
To conduct a comprehensive assessment, we employed the following objective metrics:
\begin{itemize}
    \item DNSMOS P.835\cite{reddy2022dnsmosp835nonintrusiveperceptual}: We calculate the signal quality (SIG), background quality (BAK), and overall quality (OVL) scores to evaluate speech quality.
    \item NISQA\cite{Mittag2021}: We use this metric to evaluate the naturalness and overall quality of the generated speech, with scores ranging from 1 to 5.
    \item SpeechBERT\cite{saeki2024speechbertscorereferenceawareautomaticevaluation}: We employ SpeechBERT to measure the semantic similarity and linguistic consistency between the extracted speech and the ground truth.
    \item Word Error Rate (WER): To measure content intelligibility, we calculate the WER using the robust Whisper large-v3\cite{radford2022robustspeechrecognitionlargescale} models.
    \item Speaker Similarity: We evaluate identity preservation by computing the cosine similarity of speaker embeddings extracted using WavLM \cite{Chen2022} and WeSpeaker\cite{wang2022wespeakerresearchproductionoriented}.
    \item Inference Success Rate (ISR): To quantify stability in generative streaming scenarios, we define ISR as the percentage of test samples where the model completes the autoregressive generation process. A process is invalid if the autoregressive decoding collapses due to null output. Formally, given a test set of size $N$, let $\hat{y}_{i}$ denote the generated output for the $i$-th test sample. The ISR is calculated as: 
\begin{equation}
\text{ISR} = \frac{1}{N} \sum_{i=1}^{N} \mathds{I}\left( \text{valid}(\hat{y}_i) \right) \times 100\%
\end{equation}
where $\mathds{I}(\cdot)$ is the indicator function, and $\text{valid}(\hat{y}_{i})$ is true if and only if the generation is successful.
\end{itemize}

\subsection{Baselines}
Current research in Target Speaker Extraction (TSE) is predominantly categorized into two paradigms: generative and discriminative approaches. Therefore, we benchmark our framework against two categories baselines:

\begin{itemize}
    \item \textbf{Generative Baselines:} We adopt LauraTSE\cite{b5} to benchmark our framework's streaming stability and intelligibility. LauraTSE is similar to our approach since we both use LauraGPT as the backbone. However, our approach proposes the novel Chunk-wise Interleaved Splicing Paradigm to facilitate streaming TSE. Additionally, TSELM-L\cite{tang2024tselmtargetspeakerextraction} is included as another generative baseline to provide a broader performance comparison.
    \item \textbf{Discriminative Benchmarks:} We compare with SpEx+\cite{b7} and WeSep\cite{wang2024wesepscalableflexibletoolkit}. These models serve as standard widely-adopted benchmarks in the TSE domain. While they operate with full global context (offline), comparing our streaming generative approach against these established discriminative systems allows us to assess whether our method can achieve comparable speech quality and intelligibility.
\end{itemize}

\begin{table*}[!t]
\centering

\caption{Ablation Study. Investigating the optimal input strategy for the Acoustic Refinement Language Model (ARLM) stage. Results are reported at the chunk size of 560ms. \textbf{Bold} indicates the best performance among the compared configurations.}
\label{tab:ablation_study} 

\renewcommand{\arraystretch}{1.2} 
\setlength{\tabcolsep}{5pt} 

\begin{tabular}{lcccccccc}
\toprule
\multirow{2}{*}{\textbf{Method Setup}} & \multicolumn{3}{c}{\textbf{DNSMOS$\uparrow$}} & \multirow{2}{*}{\textbf{NISQA$\uparrow$}} & \multirow{2}{*}{\textbf{SpeechBERT$\uparrow$}} & \multirow{2}{*}{\textbf{WER}$\downarrow$} & \multicolumn{2}{c}{\textbf{Sim$\uparrow$} } \\ 
\cmidrule(lr){2-4} \cmidrule(lr){8-9}
& \textbf{SIG} & \textbf{BAK} & \textbf{OVL} & & & & \textbf{WavLM} & \textbf{Wespeaker} \\
\midrule
Ref Only & 3.461 & 3.49 & 2.929 & 2.88 & 0.709 & 0.456 & 0.904 & 0.706 \\
\midrule
Ref + Sequential & \textbf{3.474} & \textbf{3.564} & \textbf{2.977} & 3.108 & 0.816 & \textbf{0.160} & \textbf{0.956} & 0.833 \\
\midrule
Ref + Interleaved & 3.473 & 3.538 & 2.963 & \textbf{3.114} & \textbf{0.817} & 0.174 & 0.955 & \textbf{0.834} \\
\bottomrule
\end{tabular}
\end{table*}

\begin{table*}[t]
\centering
\caption{Ablation study on Historical Context Refinement strategies at the chunk size of 560ms. \textbf{Bold} indicates the best performance among the compared configurations.}
\label{tab:ablation_decoder}
\setlength{\tabcolsep}{5pt}
\renewcommand{\arraystretch}{1.1}
\begin{tabular}{lcccccccc}
\toprule
\multirow{2}{*}{\textbf{Method Setup}} & \multicolumn{3}{c}{\textbf{DNSMOS$\uparrow$}} & \multirow{2}{*}{\textbf{NISQA$\uparrow$}} & \multirow{2}{*}{\textbf{SpeechBERT$\uparrow$}} & \multirow{2}{*}{\textbf{WER}$\downarrow$} & \multicolumn{2}{c}{\textbf{Sim$\uparrow$} } \\ 
\cmidrule(lr){2-4} \cmidrule(lr){8-9}
& \textbf{SIG} & \textbf{BAK} & \textbf{OVL} & & & & \textbf{WavLM} & \textbf{Wespeaker} \\
\midrule
Proposed (i.e., w/ One History Chunk) & 3.535 & 3.752 & 3.117 & \textbf{3.283} & 0.847 & 0.152  & 0.959 & 0.847 \\
w/o History Chunks & 3.473 & 3.538 & 2.963 & 3.114 & 0.817 & 0.174  & 0.955 & 0.834 \\
w/ Full History Chunks & \textbf{3.537} & \textbf{3.769} & \textbf{3.129} & 3.264 & \textbf{0.853} & \textbf{0.149}  & \textbf{0.961} & \textbf{0.848} \\
\bottomrule
\end{tabular}
\end{table*}

\section{EXPERIMENTAL RESULTS}

\subsection{Main Results: Performance Across Latency Regimes}

Table \ref{tab:main_results} presents a comparative analysis across various latency regimes. The empirical results substantiate that our framework not only enhances the performance of generative models in streaming scenario but also achieves signal quality that rivals or even surpasses established offline benchmarks.

Firstly, compared to the generative baseline (LauraTSE), our method achieves superior intelligibility and quality at low streaming latencies, followed by a decisive improvement in inference stability. At the chunk size of 560ms, our approach demonstrates a clear performance advantage, lowering the WER to 0.152 compared to the AR baseline's 0.174. This indicates that our framework better preserves semantic coherence. Furthermore, our framework ensures consistent streaming stability, maintaining a 100\% Inference Success Rate (ISR) across all settings, while generative baselines exhibit sensitivity to reduced context in low-latency scenarios. Additionally, our method at 560ms surpasses offline TSELM-L\cite{tang2024tselmtargetspeakerextraction} in signal quality (3.535 vs. 3.489) and speaker similarity (0.847 vs. 0.627). This demonstrates our framework's capability in signal quality and identity preservation.

Secondly, when benchmarked against offline discriminative models (SpEx+ and WeSep), taking a streaming input of 560ms as an example, our framework achieves performance comparable to or even better than these established offline benchmark methods. With real-time processing constraints, our method achieves a Signal Quality (SIG) score of 3.535, which is higher than both the offline SpEx+ (3.472) and WeSep (3.486). In terms of comprehensive quality, our model yields an OVL score of 3.117, which is nearly identical to the offline discriminative models WeSep (3.118) and SpEx+ (3.186).

We prioritize 560ms as the primary chunk size for analysis because it marks the critical threshold where our framework achieves 100\% stability and superior intelligibility over the AR generative baseline. This configuration satisfies funcodec’s 40ms-multiple requirement and aligns with industrial real-time standards, such as NVIDIA Riva\cite{NVIDIARivaASRCust}, which adopts 560ms as a standard for balancing low latency with high accuracy.

\subsection{Ablation Studies: Different Input Strategies for ARLM}

To determine the optimal input configuration for the Acoustic Refinement Language Model (ARLM) stage, we evaluated three distinct strategies: 
\begin{enumerate}
    \item \textbf{Ref Only:} Using only the reference embedding and predicted tokens (i.e.,$[E_{\text{ref}}, U_{\text{SELM}}^{(1:t)}]$).
    \item \textbf{Ref + Sequential:} Concatenating the reference with all mixture chunks, followed by the all sequence of predicted tokens (i.e., $[E_{\text{ref}}, C_{\text{mix}}^{(1:t)}, U_{\text{SELM}}^{(1:t)}]$).
    \item \textbf{Ref + Interleaved:} Our proposed strategy where the reference serves as a static prefix, followed by alternating mixture chunks and their corresponding predicted tokens (i.e., $[E_{\text{ref}}, C_{\text{mix}}^{(1)}, U_{\text{SELM}}^{(1)}, \dots, C_{\text{mix}}^{(t)}, U_{\text{SELM}}^{(t)}]$).
\end{enumerate}
The results are summarized in Table \ref{tab:ablation_study}.
\subsubsection{Efficacy of Mixture Embedding for ARLM stage}
We first validated the importance of incorporating the mixture context into the ARLM stage. Comparing the ``Ref Only'' strategy with the ``Ref + Sequential'' approach, we observe a substantial performance leap. At a chunk size of 560ms, incorporating the mixture context reduces WER from 0.456 to 0.160 and improves NISQA from 2.880 to 3.108. This confirms that the mixture embedding provides fine-grained acoustic details.

\subsubsection{Interleaved vs. Sequential}
When comparing the two strategies, the interleaved strategy and the sequential strategy yield comparable results, such as OVL (3.535 vs. 3.537), NISQA (3.283 vs. 3.264), WavLM (0.959 vs. 0.961) and WER (0.152 vs. 0.149). More critically, the Interleaved strategy offers a decisive advantage for real-time deployment. The Sequential approach segregates mixture and token histories, requiring new mixture chunks to be inserted before the token chunk at each step. This insertion breaks sequence contiguity, invalidates the Key-Value (KV) cache, and forces a computationally expensive re-computation of the entire history. In contrast, our Interleaved paradigm enables an efficient $O(1)$ append-only operation, where new units are simply concatenated to the existing stream. Thus, the Interleaved strategy is the superior choice, delivering optimal engineering efficiency.

\subsection{Ablation Study: Impact of Historical Context Refinement}

A critical challenge in chunk-wise streaming generation is the boundary discontinuity between adjacent segments, which often disrupts phase continuity and semantic coherence. To mitigate this, we evaluate Historical Context Refinement mechanism that reuses the refined embedding from the step $t-1$ chunk as prompt. As demonstrated in Table \ref{tab:ablation_decoder}, omitting this mechanism (w/o History Chunks) results in a substantial performance decline, where the WER rises from 0.152 to 0.174 and the NISQA score drops from 3.283 to 3.114. These findings empirically substantiate that this mechanism functions as a refinement module, significantly improving speech quality and intelligibility in low-latency scenarios.

We further examine the trade-off between local and global history by comparing our proposed One History Chunk strategy with the Full History Chunks strategy. While the Full History Chunks approach yields marginal improvements, such as further reducing the WER from 0.152 to 0.149 and increasing the BAK score from 3.752 to 3.769, the improvements in perceived quality are minimal and have effectively reached a saturation point. From an engineering standpoint, while preserving Full History Chunks yields marginal performance gains, it introduces cumulative computational overhead and memory consumption, such a scaling characteristic is inherently incompatible with the stringent constraints of real-time streaming deployment. Consequently, we adopt the Last Context Refinement as the optimal configuration, operating in an append-only manner during inference to ensure the balance between speech quality and processing efficiency.

\subsection{Real-Time Factor (RTF) Analysis}
To assess the computational efficiency of our proposed framework for real-world deployment, we evaluated the Real-Time Factor (RTF) across different GPU architectures:

\begin{equation}
\text{RTF} = \frac{T_{\text{proc}}}{T_{\text{speech}}}
\end{equation}
where $T_{\text{proc}}$ denotes the end-to-end time to process a speech (include feature extraction, model inference, and waveform reconstruction), and $T_{\text{speech}}$ represents the duration of the speech. An RTF less than 1.0 indicates that the system can process speech faster than real-time, which is a prerequisite for online streaming. We conducted the evaluation using the chunk size of 560ms. As shown in Table \ref{tab:rtf_analysis}, our model achieves an RTF significantly lower than 1.0 across all tested platforms:
\begin{itemize}
    \item On the NVIDIA V100, a widely used training GPU, the RTF is 0.433, confirming that even older generation hardware can support our streaming inference comfortably.
    \item On the consumer-grade NVIDIA RTX 4090, the RTF drops to 0.248, demonstrating high efficiency for edge or workstation deployment.
    \item On the NVIDIA L40S, the model achieves an impressive RTF of 0.182. This indicates the system utilizes less than 20\% of the computational capacity for real-time processing, leaving ample headroom for downstream tasks.
\end{itemize}

\begin{table}[t]
\centering
\caption{Real-Time Factor (RTF) Analysis. Measured at a chunk size of 560ms across different hardware platforms. Lower RTF is better.}
\label{tab:rtf_analysis}
\setlength{\tabcolsep}{10pt}
\begin{tabular}{l|c}
\toprule
\textbf{Hardware Platform} & \textbf{RTF$\downarrow$} \\ \hline
NVIDIA V100 & 0.433 \\
NVIDIA RTX 4090 & 0.248 \\
NVIDIA L40S & 0.182 \\ 
\bottomrule
\end{tabular}
\end{table}

\section{Conclusion}
This work presents the first autoregressive (AR) generative backbone tailored for streaming Target Speaker Extraction, filling a critical research void. By introducing the Chunk-wise Interleaved Splicing Paradigm, we enable streaming Target Speaker Extraction while maintaining causality and efficient append-only operation, decisively eliminating the stability collapse seen in baseline AR generative models. The historical context refinement significantly enhances speech quality and intelligibility by leveraging historical information. Empirical results demonstrate superior intelligibility, 100\% stability in low latency compare to baseline AR generative model, and can match or surpass offline baseline models. In addition, we achieved a Real-Time Factor (RTF) of 0.248 on consumer-grade GPU, which is the key metric for streaming scenarios. Future efforts will focus on improving speaker similarity and speech quality under ultra-low latency constraints.

\bibliographystyle{IEEEbib}
\bibliography{references}

\end{document}